\documentclass[amsmath,amssymb,nofootinbib]{revtex4}

\usepackage{bbold}
\usepackage{slashed}
\usepackage{bigints}

\begin{document}

\title{Existence of Matter as a Proof of the Existence of Gravitational Torsion}

\author{Carl F. Diether III\footnote{\!\!Corresponding author}}
 \email{fred.diether@einstein-physics.org}

\author{Joy Christian}
 \email{joy.christian@einstein-physics.org}

\affiliation{Einstein Centre for Local-Realistic Physics, 15 Thackley End, Oxford OX2 6LB, United Kingdom}

\maketitle

\begin{center}
\bf{Abstract}
\end{center}

\begingroup
\addtolength\leftmargini{0.2in}
\begin{quote}
One of the reasons why Einstein-Cartan-Sciama-Kibble theory --- which includes torsion due to the intrinsic spin of matter --- is not widely accepted as a viable theory of gravity, is the lack of any experimental evidence for the gravitational torsion. But at the heart of all matter are elementary\break fermions, which are mathematically described as spinors, and can be viewed as defects in spacetime associated with torsion. Therefore, in this essay we argue that the existence of matter itself provides a proof that gravitational torsion exists, albeit only inside of matter, and hence it does not propagate.  
\end{quote}
\endgroup

\vfill

\vfill

\vfill

\vfill

\vfill

\vfill

\vfill

\vfill

\vfill

\begin{center}
\it{Essay written for the Gravity Research Foundation 2019 Awards for Essays on Gravitation.}
\end{center}

\vfill\eject

\parskip 5pt

\baselineskip 13.1pt

\section{Introduction}

For over a century Einstein's theory of gravity has provided remarkably accurate and precise predictions for the behaviour of macroscopic bodies within our universe. For the elementary particles in the quantum realm, however, Einstein-Cartan theory of gravity may be more appropriate, because it incorporates spinors and associated torsion within a covariant description \cite{Hehl1976}\cite{Trautman}. For this reason there has been considerable interest in Einstein-Cartan theory, in the light of the field equations proposed by Sciama \cite{Sciama} and Kibble \cite{Kibble}. For example, in a series of papers Poplawski has argued that Einstein-Cartan-Sciama-Kibble (ECSK) theory of gravity \cite{Hehl-Datta} solves many longstanding problems in physics \cite{Poplawski-1}\cite{Poplawski-2}\cite{Poplawski-3}\cite{Poplawski-4}\cite{Poplawski-5}\cite{Poplawski-6}. His concern has been to avoid singularities endemic in general relativity by proposing that our observed universe is perhaps a black hole within a larger universe \cite{Poplawski-2}. However, so far ECSK theory has not enjoyed universal acceptance as a viable theory of gravity, partly because there has been no experimental evidence supporting the existence of gravitational torsion. Moreover, even though theoretical arguments in support of gravitational torsion are compelling (for a recent review see \cite{Cabral} and references therein), it is quite difficult to test for its existence in either terrestrial or cosmological experiments. As a result, the role of torsion in a fundamental theory of gravity has remained subject to individual preferences and interpretations. With that in mind, we propose an interpretation that the mere existence of matter {\it is} the proof that gravitational torsion due to the intrinsic spin of elementary fermions exists. 

To that end, for an intuitive understanding of what we mean by torsion in this context, consider a rigid rod which can be bent. If that corresponds to the curvature of spacetime due to energy momentum tensor, then torsion would be a twist of the rod's mechanical structure. We therefore assume that if spacetime can be ``bent", then it can also be twisted. And, consequently, it can be both ``bent" and ``twisted" at the same time. But spacetime twisted in this manner will induce a ``defect", amounting to a non-closure of a loop in the corresponding manifold.

\section{If Matter Exists, then Gravitational Torsion Exists}

In Ref.~\cite{diether} we have presented an argument for the necessity of gravitational torsion for canceling out the self-energy stemming from the interaction of a charge with the field that the charge itself creates within a charged elementary fermion, to produce the rest mass-energy of the elementary fermions. This resulted in discovering that the radii of elementary fermions is of the order of $10^{-33}$ m, very close to Planck length.  Our discovery is based on the Hehl-Datta equation from their 1971 paper \cite{Hehl-Datta}, which, in an alternate form, can be written as
\begin{equation}
i\hbar\,\gamma^k\psi_{:k}-\frac{3\kappa\hbar^2c}{8}\left(\bar{\psi}\gamma^5\gamma_k\psi\right)\gamma^5\gamma^k\psi = mc \, \psi. \label{HD}
\end{equation}
Here $_{:k}$ represents the covariant derivative and $\kappa = 8\pi G/c^4$.  We will call the second term on the left hand side of eq.~(\ref{HD}) the Hehl-Datta term, which is the result of the torsion translation to a spin-squared momentum.  Thus the complete Quantum Electrodynamics (QED) Lagrangian density for ECSK theory in units of $\hbar = c = 1$ is given by 
\begin{equation}
\mathfrak{L}_{\rm QED} =  \sqrt{-g\,}\left\{i\,\bar{\psi}\gamma^k\partial_k \psi+e\,\bar{\psi}\gamma^k A_k \psi-\frac{3\kappa}{16}\left(\bar{\psi}\gamma^5\gamma_k\psi\right)\bar{\psi}\gamma^5\gamma^k\psi - m \,\bar{\psi} \psi-\frac{1}{4} F_{k l} F^{k l}\right\}. \label{QED1}
\end{equation}
Here $e$ is the charge of the positron and $m$ is its mass.  While this Lagrangian is based on a minimal coupling, we consider it to be a more complete Lagrangian for QED in curved spacetime because it contains the Hehl-Datta term.

The result of our investigation using S-Matrix analysis for the rest frame of a charged lepton is the following formula after renormalization of the electrostatic term:
\begin{equation}
\frac{3\,\alpha \hbar c}{8\pi r} - \frac{\,3\kappa (\hbar c)^2}{16\,r^3} = m c^2. \label{result}
\end{equation}
The first term in this equation is the energy due to electrostatic self-energy. The second term is from the Hehl-Datta term that represents the mechanical gravitational torsion energy due to the intrinsic spin interacting with itself. The remaining terms in the Lagrangian do not affect the rest frame result.  It is easy to see from the above expression that, given the rest mass energy for the charged lepton, one can solve for the radius, $r$, of the charge distribution. Since the Hehl-Datta term is a cubic term, one obtains two positive solutions for $r$: One of them very close to Planck length, and another one of order of the reduced Compton wavelength, because for the Hehl-Datta term effectively zero we have $r= 3\alpha \hbar /(8\pi m c)=3\alpha \lambda_C/8\pi$, where $\lambda_C$ is the reduced Compton wavelength. This puts the second radius near the classical electron radius.  For an electron, we find the two radii to be $\approx 1.188 \times 10^{-33}\,\rm{m}$ and $3.364\times 10^{-16}\,\rm{m}$, respectively. Here the appearance of the gravitational coupling, $G$, in a Dirac-type equation may seem surprising. But by equating ${G\,\hbar^2 c/c^4}$ with ${\hbar\, l_P^2}$ we can see that the coupling involves a square of Planck length where part of the spin-squared is absorbed into the Planck length coupling along with the gravitational constant. If we now set the mass in the above formula to zero, then we arrive at a simple constant for which the two energies cancel out completely:
\begin{equation}
\boxed{r_t = {\frac{2\pi}{\sqrt{\alpha\,}}}\;l_P}
\end{equation}
The numerical value of this radius works out to be
\begin{equation}
r_t \approx 1.189 \times 10^{-33}\, \rm{m}\,.
\end{equation}
This is our ``cancellation radius" --- {\it i.e.}, the radius of the charge distribution at which the electromagnetic self-energy of the fermion and its gravitational torsion energy (which stems from the self-interaction its intrinsic spin) cancel each other out completely. Evidently, it is a simple function of only Planck length ${l_P}$ and the fine structure constant ${\alpha}$.

Returning to our main goal, we now ask: How would matter behave if gravitational torsion was not helping it to produce the observed rest mass-energy? Undoubtedly, provided the fermions' radii had a cutoff near Planck length, they would correspond to huge rest mass-energy near the Planck mass. Consequently, our universe would be quite different, if it existed at all. To be sure, this difficulty is traditionally overcome with the renormalization procedure, despite recurrent skepticism about its use \cite{diether}. By contrast, we believe that the existence of matter in the universe itself provides a proof that gravitational torsion exists, albeit only inside of matter, and hence it fails to propagate.  

\section{Torsion Induces a Defect within $V_4$ Spacetime}

To motivate our conjecture that matter is a ``defect" within spacetime, we quote from the book ``Geometric Algebra and Applications to Physics", authored by Sabbata and Datta \cite{sabbata}: 
\begin{quote}\ldots but we will emphasize the fundamental fact that one of the most important geometrical properties of torsion is that a closed contour in an $U_4$ manifold becomes, in general, a nonclosed contour in the flat space-time $V_4$.  This nonclosure property, that is, the fact that the integral
\begin{equation}
I^{\alpha} = \oint Q_{\beta\gamma}^{\alpha}\, dS^{\beta\gamma} \neq 0 \notag %\label{sabbata}
\end{equation}
(where $dS^{\beta\gamma}= dx^{\beta}\wedge dx^{\gamma}$ is the area element enclosed by the loop) over a closed infinitesimal contour is different from zero, can be treated as defects in space-time in analogy to the geometrical description of dislocations (defects) in crystals; this can constitute a way to move toward the quantization of gravity, which means quantization of space-time itself.
\end{quote}
Here ${U_4}$ is Einstein-Cartan spacetime, ${V_4}$ is general relativistic spacetime, $Q$ is the torsion tensor, and $I$ has dimensions of length.  Since torsion is entirely confined to matter \cite{Hehl1976}, its interpretation is quite straightforward: Matter itself is a quantization of spacetime as well as a defect within $V_4$ spacetime.  The torsion is confined to matter because it is generated by the intrinsic spin of that matter which interacts with itself. This suggests two types of quantization of gravity. The graviton resulting from a quantization of the usual curvature of the ${V_4}$ spacetime, and matter resulting from the torsion within ${U_4}$. We also suspect that there may be some kind of rich symmetry breaking associated with matter beyond the simple fact that $V_4$ spacetime symmetry is broken by torsion and becomes a $U_4$ symmetry.

In summary, what we are proposing is that all of spacetime external to matter remains a $V_4$ manifold, whereas matter ({\it i.e.}, elementary fermions) constitutes a $U_4$ manifold. Now, one sometimes comes across an objection to ECSK theory of gravity, which has been spelled out in Ref.~\cite{Hehl2007} from the textbook by Ohanian and Ruffini \cite{ohanian}:
\begin{quote}
    If $\Gamma^\beta_{\nu\mu}$ [affine connection] were not symmetric, the parallelogram would fail to close.  This would mean that the geometry of the curved spacetime differs from a flat geometry even on a small scale - the curved spacetime would not be approximated locally by a flat spacetime.
\end{quote}
But Hehl and Obukhov \cite{Hehl2007} have argued that this objection is not valid because ``the Riemann-Cartan geometry [of ${U_4}$] is Euclidean `in the infinitesimal'." However, in our view the geometry of spacetime {\it does} differ from the Euclidean ``in the infinitesimal" because we may have a $U_4$ fermion at the ``defect" or the non-closure point of the ${V_4}$ manifold. Thus a fermion is the result of the defect. This is simpler explanation compared to that offered by Hehl and Obukhov.

\section{Discussion}

The ECSK theory of gravity extends general relativity to include the intrinsic spin of matter, with fermionic fields such as those of quarks and leptons providing natural sources of torsion. Torsion, in turn, modifies the Dirac equation for elementary fermions by adding to it a cubic term in the spinor fields, as observed, for example, by Kibble, Hehl, and Datta \cite{Hehl1976}\cite{Kibble}\cite{Hehl-Datta}. However, torsion is somewhat hidden in the Hehl-Datta equation, because it enters the Dirac equation algebraically. But we can demonstrate that it is indeed there by inserting the energy density version of the Hehl-Datta term into Einstein's field equations for our cancellation radius: 
\begin{equation}
G^{00}_{spin}= -\frac{8 \pi  G }{ c^4} \frac{3\, \pi  l_P \hbar c}{2 \, r_t^6} \approx -2.863 \times 10^{60}\;m^{-2}.
\end{equation}
Note that gravitational torsion is negative relative to curvature.

Within the ECSK theory gravitational effects near micro scales are not necessarily weak. On the other hand, since torsion is produced in the ECSK theory by the spin density of matter, it is confined to that matter, and thus is a very short range effect, unlike the infinite range effect of Einstein's gravity produced by mass-energy.

In our view, any generalization of general relativity should include both torsion and teleparallelism, because, as explained by Hehl and Obukhov \cite{Hehl2007}, torsion arises from the translational gauge invariance. In fact, it is simply a translation gauge field strength. Moreover, ECSK theory gives rise to a new spin-spin contact self-interaction, which can be measured, in principle, by the precession of elementary particle spins in torsion fields according to Hehl and Obukhov.  However, it seems unlikely that gravitational torsion can be measured directly in this manner, because, as we have shown in \cite{diether}, the torsion fields are entirely confined within the elementary fermion and are completely compensated for by the electrostatic type fields. Moreover, any uncompensated torsion fields would have anti-gravity properties which would have been observed in experiments. However, within our interpretation, torsion {\it can} be measured {\it indirectly}, by simply measuring the rest mass of an elementary fermion. 

In Ref.~\cite{Hehl2007} Hehl and Obukhov ask: ``How can a local observer at a point P with coordinates $x^i$ tell whether his or her space carries torsion and/or curvature?" Within our interpretation the answer to this question is quite simple. If there is matter at point P then there is gravitational torsion within the matter, in addition to curvature.  On the other hand, if there is no matter at point P, then there could be some amount of curvature depending on the matter/energy near the point P, but there will be no gravitational torsion in that case.

It is widely accepted that in the standard model of particle physics charged elementary fermions acquire masses via the Higgs mechanism.  Within this mechanism, however, there is no satisfactory explanation for how the different couplings required for the fermions are produced to give the correct values of their masses. While the Higgs mechanism does bestow masses correctly to the heavy gauge bosons and a massless photon, and while our result (\ref{result}) above does not furnish a fundamental explanation for the fermion masses either, what we have proposed in this essay --- namely, that existence of matter itself is a proof of the existence of gravitational torsion --- seems worthy of further research.

\end{document}